\documentclass[twoside,twocolumn,a4paper,9pt]{article}
\usepackage{extsizes}
\usepackage[super,sort&compress,comma]{natbib} 
\usepackage[version=3]{mhchem}
\usepackage[left=1.5cm, right=1.5cm, top=1.785cm, bottom=2.0cm]{geometry}
\usepackage{balance}
\usepackage{mathptmx}
\usepackage{sectsty}
\usepackage{graphicx} 
\usepackage{lastpage}
\usepackage[format=plain,justification=justified,singlelinecheck=false,font={stretch=1.125,small,sf},labelfont=bf,labelsep=space]{caption}
\usepackage{float}
\usepackage{fancyhdr}
\usepackage{fnpos}
\usepackage[english]{babel}
\addto{\captionsenglish}{%
  
}
\usepackage{array}
\usepackage{droidsans}
\usepackage{charter}
\usepackage[T1]{fontenc}
\usepackage[usenames,dvipsnames]{xcolor}
\usepackage{setspace}
\usepackage[compact]{titlesec}
\usepackage{hyperref}

\definecolor{cream}{RGB}{222,217,201}

\begin{document}

\pagestyle{fancy}
\thispagestyle{plain}
\fancypagestyle{plain}{
\renewcommand{\headrulewidth}{0pt}
}

\makeFNbottom
\makeatletter
\renewcommand\LARGE{\@setfontsize\LARGE{15pt}{17}}
\renewcommand\Large{\@setfontsize\Large{12pt}{14}}
\renewcommand\large{\@setfontsize\large{10pt}{12}}
\renewcommand\footnotesize{\@setfontsize\footnotesize{7pt}{10}}
\makeatother

\renewcommand{\thefootnote}{\fnsymbol{footnote}}
\renewcommand\footnoterule{\vspace*{1pt}%
\color{cream}\hrule width 3.5in height 0.4pt \color{black}\vspace*{5pt}} 
\setcounter{secnumdepth}{5}

\makeatletter 
\renewcommand\@biblabel[1]{#1}            
\renewcommand\@makefntext[1]%
{\noindent\makebox[0pt][r]{\@thefnmark\,}#1}
\makeatother 
\renewcommand{\figurename}{\small{Fig.}~}
\sectionfont{\sffamily\Large}
\subsectionfont{\normalsize}
\subsubsectionfont{\bf}
\setstretch{1.125} 
\setlength{\skip\footins}{0.8cm}
\setlength{\footnotesep}{0.25cm}
\setlength{\jot}{10pt}
\titlespacing*{\section}{0pt}{4pt}{4pt}
\titlespacing*{\subsection}{0pt}{15pt}{1pt}

\fancyfoot{}
\fancyfoot[LO,RE]{\vspace{-7.1pt}\includegraphics[height=9pt]{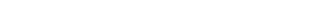}}
\fancyfoot[CO]{\vspace{-7.1pt}\hspace{13.2cm}\includegraphics{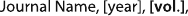}}
\fancyfoot[CE]{\vspace{-7.2pt}\hspace{-14.2cm}\includegraphics{head_foot/RF}}
\fancyfoot[RO]{\footnotesize{\sffamily{1--\pageref{LastPage} ~\textbar  \hspace{2pt}\thepage}}}
\fancyfoot[LE]{\footnotesize{\sffamily{\thepage~\textbar\hspace{3.45cm} 1--\pageref{LastPage}}}}
\fancyhead{}
\renewcommand{\headrulewidth}{0pt} 
\renewcommand{\footrulewidth}{0pt}
\setlength{\arrayrulewidth}{1pt}
\setlength{\columnsep}{6.5mm}
\setlength\bibsep{1pt}

\makeatletter 
\newlength{\figrulesep} 
\setlength{\figrulesep}{0.5\textfloatsep} 

\newcommand{\topfigrule}{\vspace*{-1pt}%
\noindent{\color{cream}\rule[-\figrulesep]{\columnwidth}{1.5pt}} }

\newcommand{\botfigrule}{\vspace*{-2pt}%
\noindent{\color{cream}\rule[\figrulesep]{\columnwidth}{1.5pt}} }

\newcommand{\dblfigrule}{\vspace*{-1pt}%
\noindent{\color{cream}\rule[-\figrulesep]{\textwidth}{1.5pt}} }

\makeatother

\twocolumn[
  \begin{@twocolumnfalse}
{\includegraphics[height=30pt]{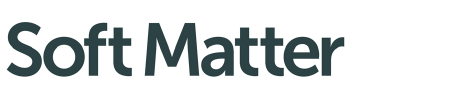}\hfill\raisebox{0pt}[0pt][0pt]{\includegraphics[height=55pt]{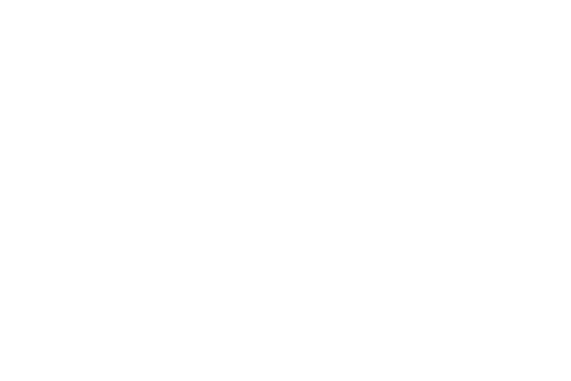}}\\[1ex]
\includegraphics[width=18.5cm]{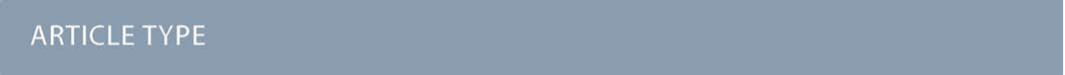}}\par
\vspace{1em}
\sffamily
\begin{tabular}{m{4.5cm} p{13.5cm} }

\includegraphics{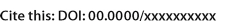} & \noindent\LARGE{\textbf{Phase behaviour of coarse-grained fluids}} \\
\vspace{0.3cm} & \vspace{0.3cm} \\

 & \noindent\large{V.~P.~Sokhan,\textit{$^{a\ast}$} M.~A.~Seaton,\textit{$^{a\ddag}$} and I.~T.~Todorov\textit{$^{a}$}} \\

\includegraphics{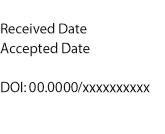} & \noindent\normalsize{Soft condensed matter structures often challenge us with complex many-body phenomena governed by collective modes spanning wide spatial and temporal domains. In order to successfully tackle such problems, mesoscopic coarse-grained (CG) statistical models are being developed, providing a dramatic reduction in computational complexity.
CG models provide an intermediate step in the complex statistical framework of linking the thermodynamics of condensed phases with the properties of their constituent atoms and molecules. These allow us to offload part of the problem to the CG model itself and reformulate the remainder in terms of reduced CG phase space. However, such exchange of pawns to chess pieces, or `Hamiltonian renormalization', is a radical step and the thermodynamics of the primary atomic and CG models could be quite distinct. 
Here, we present a comprehensive study of the phase diagram including binodal and interfacial properties of a dissipative particle dynamics (DPD) model, extended to include finite-range attraction to support the liquid-gas equilibrium.
Despite the similarities with the atomic model potentials, its phase envelope is markedly different featuring several anomalies such as an unusually broad liquid range, change in concavity of the liquid coexistence branch with variation of the model parameters, volume contraction on fusion, temperature of maximum density in the liquid phase and negative thermal expansion in the solid phase.
These results provide new insight into the connection between simple potential models and complex emergent condensed matter phenomena.} \\

\end{tabular}

 \end{@twocolumnfalse} \vspace{0.6cm}
  ]

\renewcommand*\rmdefault{bch}\normalfont\upshape
\rmfamily
\section*{}
\vspace{-1cm}

\footnotetext{\textit{$^{a}$~Scientific Computing Department, Science and Technology Facilities Council, STFC Daresbury Laboratory, Sci-Tech Daresbury, Keckwick Lane, Daresbury, Cheshire WA4 4AD, UK}}
\footnotetext{$^{\ast}$~E-mail: vlad.sokhan@stfc.ac.uk}
\footnotetext{$^{\ddag}$~E-mail: michael.seaton@stfc.ac.uk}

\section*{Introduction}
Coarse-graining (CG) is now well established as an essential ingredient of multiscale modelling frameworks\cite{Noid08} and provides accelerated pathways to characterizing soft condensed matter\cite{Nage17} including complex fluids,\cite{Pete09} biological membranes\cite{Sade20} and ionic solutions.\cite{Vazq20} Despite the inevitable loss in structural detail at the CG level, its enormous potential in expanding the length and time scales to address large collective phenomena drives current advances in complex condensed phases characterization. Soft matter often contains colloids and polymer solutions\cite{Liko01} and in order to achieve the desired speedup in computer simulation of such systems both the solutes and solvents need to be coarse-grained congruently. Most forthrightly, CG can be carried out in biological and polymer systems by mapping (macro-)molecular fragments and moieties of chemically connected atomic fragments to `beads' interacting via CG forces.\cite{Alle17,Saun13} Here, CG methods have scored rapid successes from their arrival in the late 1960s, culminating in the 2013 Nobel Prize in Chemistry awarded to Martin Karplus, Michael Levitt and Arieh Warshel `for the development of multiscale models for complex chemical systems'.\cite{Karp14} Their ideas on simplified protein structures\cite{Levi75} are now broadly adopted in biomolecular simulation and software, e.g., in the Martini force field.\cite{Souz21}

Conversely, CG mapping of solvents, and fluid phases in general, is a less clear-cut problem\cite{Erik09,Espa17,Han18} apparently due to the very problem of consistent definition of a fluid element (aka `blob') in statistical mechanics. Unrestricted motion of atoms in the fluid phases stipulates an open system view of CG particles\cite{Cicc19} and approaches based on Voronoi tessellation,\cite{Serr01,Flek99,Han18} iterative $k$-means clustering of MacQueen,\cite{MacQ67,Hadl10} ghost `blob' microprobes,\cite{Ayton04} and Brownian Quasiparticles\cite{Sokh19} have been put forward to tackle this matter, although the problem of an `inter-blob' potential still remains largely open and no closed form solution exists so far.\cite{Espa17} Crucially, although CG interactions can be cast in the form of an effective pair potential, its functional form is likely to be different from that of atomistic systems.\cite{Liko01,Hans02} Accuracy of CG mapping depends critically on the functional form of the ansatz, which could only be hypothesised. This already poses a problem with bonded interactions,\cite{Liko01,Flor50,Hans02} but presents a fairly non-trivial problem for non-bonded interactions that involves genuine perception.\cite{Espa17} Existing methods provide the means to optimise the parameters of the model, but not its functional form. Common ans{\"a}tze validated at the atomic scale, such as the omnipresent Lennard-Jones (LJ) potential, are not necessarily the best choices for generic effective CG potentials,\cite{Bern72} where a soft Gaussian would be the likely candidate. In fact, the utility of the LJ potential has been recently questioned even at the atomic scale.\cite{Wang20} Furthermore, although it usually goes unquestioned in coarse-graining, various bottom-up approaches\cite{Izve05,Shel08} hold the underlying atomic representation as a ground-truth structure. It is important to realise, however, that the atomic data are themselves based on effective empirical pair potentials, liable to be inaccurate away from their training sets.

For large-scale fluid dynamics problems dissipative particle dynamics (DPD)\cite{Groo97} is arguably the most widely used particle-based method describing various aspects of multiphase flow at the mesoscale.\cite{Espa17} Being generic, fast and simple to implement, DPD expands the range of accessible time and spatial scales by several orders of magnitude.\cite{Groo01} DPD introduces a remarkably simple, albeit fundamentally challengeable ansatz: harmonic repulsion between the beads representing fluid elements. The model has been instructive in broad range of applications including colloids, polymers, fluid mixture and amphiphilic systems.\cite{Espa17} However, employing only soft repulsive forces has a serious drawback -- DPD covers both fluid and solid phases, but misses gas-liquid coexistence.\cite{Groo08} In other words, it is incapable of capturing such vital liquid surface phenomena as wetting, spreading, and many important capillary effects.\cite{Rowl82a,Ghou16} Attraction is a prerequisite for a stable liquid-gas interface,\cite{Hemm76} with different requirements for systems of different dimensionality. The relative range of attraction in CG potentials is likely be reduced,\cite{Teje94} though it is believed that for 3D systems any finite-range attraction guarantees a first-order liquid-gas transition.\cite{Hemm70} 

A currently adopted way to remedy this serious drawback in the DPD model is to add attraction to its conservative forces by including many-body interactions through an ad hoc density-dependent term, as done in many-body DPD (MB-DPD),\cite{Warr03} with a caveat of thermodynamic inconsistency.\cite{Loui02} However, explicit many-body terms are not strictly necessary for gas-liquid coexistence, and simple fluid models that include many-body terms only effectively are able to capture a wealth of capillary phenomena within the pairwise approximation.\cite{Rowl82a}
In a more straightforward approach the attraction can be included by extending the DPD model beyond the harmonic approximation and combining two terms of different powers: a positive term of higher power would give an anharmonic repulsion, and a lower-power term taken with a negative sign would provide the attraction. For example, Groot and Stoyanov \cite{Groo08} used a quadratic force in their sticky elastic sphere model to study colloidal dispersions. The standard DPD model specifies the sign of the weight function but not its functional form,\cite{Groo97} and lifting the requirement of nonnegativity one can explore this freedom in full. Thus, combining two terms with opposite signs is equivalent to a single term with a weight function that is negative within a certain domain. Furthermore, since the conservative and Langevin weight functions of the model are not generally correlated, the latter can be preserved, retaining the hydrodynamics of the original DPD. 
Here, we explore this approach to extend the DPD model to all three phases of matter and relate the corresponding thermodynamics at meso- and nanoscales.

In the van der Waals (vdW) picture of simple liquids, the short-range structure and correlations are defined primarily by inter-particle repulsion, while attraction plays only a minor role.\cite{Week71} However, when it comes to thermodynamic properties, both attraction and repulsion actively shape the phase diagram,\cite{Smit91} which is liable to differ for the CG system as the result of entropy loss. Even for `bottom-up' derived CG potentials for bonded species,\cite{Noid13} isomorphism between atomic and CG phase envelopes is not guaranteed.
To establish a corresponding connection for systems with coarse-grained fluid particles reliable effective potentials are required. Although their rigorous form is not available, a clue may be found from polymer solutions by taking the view of a fluid blob as a limiting case of a star or ring polymer with the force constant between the monomers $k\to 0$.
Both theory\cite{Flor50} and simulation \cite{Liko01} provide evidence that the effective potential between the polymers is soft, repulsive and short-ranged. Thus, it is well known that the interactions between the polymer chains in solution can be described by a Gaussian.\cite{Flor50} Similarly, ring polymers are coarse-grained to ultrasoft bounded potentials.\cite{Narr10} Star polymers are also shown to be exponentially decaying outside their corona diameter.\cite{Liko98} Thus one could conclude that the effective potential between the fluid blobs is also bounded and short-ranged. Softness of the CG potential for non-bonded interactions has also been demonstrated from the atomistic perspective,\cite{Klap04} and moreover, the DPD method is entirely hinged on this approach.

Simple potentials do not necessarily lead to simple thermodynamics, which could be rich in unusual and anomalous features.\cite{Male07}
There are several soft potentials including the Stell-Hemmer\cite{Hemm70,Hemm76} and Jagda models\cite{Jagla98} which are known for several anomalies.\cite{Sadr98,Wild02,Barr06} Their anomalies are ascribed to the presence of regions with negative curvature in the potentials (`core-softened' potentials). On the other hand, our extension of the DPD model eschews core-softening and its analytic potential remains convex throughout its entire range, being likely the first of this type to exhibit a full set of anomalies.

The principal objective of the present paper is to examine the phase behaviour of a generic model for the coarse-grained solvent, contrasting it with that of a fully fledged atomistic solvent. Often, the origins of complex thermodynamic phenomena are subtly encoded in the details of interparticle interactions. However, a simple general model could help revealing the underlying physical mechanisms. 
In the next section we outline the generalised potential, termed $n$DPD, and provide details of simulations. We then present the results of condensed phase simulations using $n$DPD potentials, demonstrating stable liquid and solid phases for all three studied orders, $n=2,3,4$. We further discuss the observed anomalies in the thermodynamic properties and outline our conclusions.

\begin{figure}
\centering
\includegraphics[width=.8\linewidth]{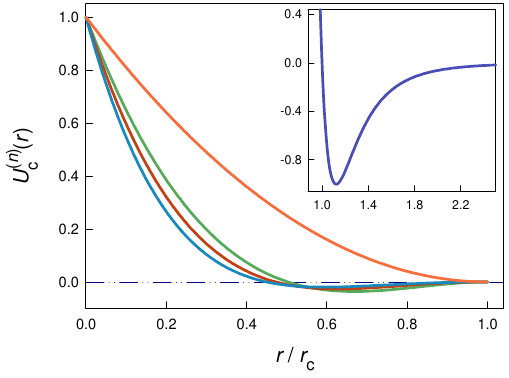}
\caption{Standard DPD (orange line) and proposed $n$DPD interaction model (green, red, and blue lines for $n$ = 2, 3, and 4, respectively) showing the attractive region in $r$ from $\sigma$ $(\sim 0.5r_{\text{c}})$ up to the characteristic length scale (cutoff distance) $r_{\text{c}}$. In the inset, the standard Lennard-Jones potential is shown in the LJ reduced units, $\varepsilon=\sigma=1$.} 
\label{fig:one}
\end{figure}

\section*{Simulation methods}
\subsection*{DPD interactions with attraction}
In classical statistical simulations an interaction between spherically-symmetric species is well described by the Mie\cite{Mie03} potential, which in the current notation is written as 
\begin{equation}
  U^{(n,m)}(r_{ij}) = \left(\frac{n}{n-m}\right)\left(\frac{n}{m}\right)^{\frac{m}{n-m}}\varepsilon\left[\left(\frac{\sigma}{r_{ij}}\right)^n
       - \left(\frac{\sigma}{r_{ij}}\right)^m\right],
\end{equation}
where $\varepsilon$ and $\sigma$ are the parameters defining the energy and length scales correspondingly, and $n$ and $m$ ($n>m>3$) are the exponents defining together the potential well width and the steepness of the potential at short range. The case $n=12$, $m=6$ is the archetypal Lennard-Jones (LJ) potential,\cite{Alle17} illustrated in the inset to Fig.~\ref{fig:one}. Although this generic potential is able to describe quantitatively a wide range of classical atomistic systems including their liquid phases and critical phenomena, its use in CG modelling is hindered by several factors. Firstly, it is too steep at short range\cite{Liko01} and diverges at the origin. Furthermore, as a potential of infinite range, it requires truncation in numerical schemes,\cite{Wang20} and the thermodynamics of a truncated potential are sensitive to the details of truncation.\cite{Cous95}
The last two issues have been recently discussed from the atomic perspective\cite{Wang20} but the proposed solution, based on potentials diverging at the origin, is again aimed at atomistic modelling or coarse-graining bonded potentials.

Unlike CG potentials for chemically bonded species, diverging at the origin, coarse-grained non-bonded interactions in complex fluids are likely to be `soft', i.e., bound everywhere,\cite{Hans02,Liko98,Liko01} which is a consequence of intra-particle motion.\cite{Liko98,Klap04} Perhaps the most general form of the soft `overlap' potential is the Gaussian potential, which has re-appeared in classical statistical studies many times in different guises.\cite{Flor50,Bern72,Stil76,Liko98,Purt10} However, its usage has one serious shortcoming: a vanishing force at the origin. A bounded potential is also at the crux of the DPD model \textendash{} there, the force is at its maximum at the origin, and its softness and continuity enables fast local equilibration and exploration of phase space.

A possible way to include both repulsion and attraction in the DPD framework is based on the cubic spline weight functions of smoothed particle hydrodynamics (SPH).\cite{Liu06} Such ad hoc solutions would nevertheless provide little generality and insight, and as with the case of Gaussian potential, would lead to a vanishing force at the origin.
Instead, we put forward a new potential class for CG simulations building on the DPD model as the lowest term in a series expansion, and construct the CG potential by combining two different-order terms: the higher order, $n$, describing the repulsion and the lower order, $m$, describing the attraction and taken with the negative sign.
We conjecture that by augmenting the DPD potential with a short-ranged attractive part we retain the DPD advantages of accurately describing the collective dynamics of fluid phases, adding the benefit of including the rich world of capillary phenomena due to integrated positive surface tension.
We start from the DPD model, where there are two types of forces between the particles (beads): the conservative (potential) force, which defines the thermodynamic state of the substance (fluid), and the Langevin (dissipative and random) forces, which are responsible for its hydrodynamic behaviour. Here, we focus on the first type, $\mathbf{F}_\text{c}(r_{ij})$, which in the standard DPD model decreases linearly with the interparticle separation $r_{ij}\equiv\vert\mathbf{r_{ij}}\vert$,
\begin{equation}
\mathbf{F}_\text{c}(r_{ij}) = A_{ij}\left(1-\frac{r_{ij}}{r_\text{c}}\right)\mathbf{e}_{ij},\label{eqn:dpdf}
\end{equation}
giving rise to a quadratic potential
\begin{equation}
U(r_{ij}) = \frac{1}{2}A_{ij}r_\text{c}\left(1-\frac{r_{ij}}{r_\text{c}}\right)^2.
\end{equation}
Here, $\mathbf{e}_{ij}\equiv\mathbf{r}_{ij}/r_{ij}$ is the unit vector in the interparticle direction and $r_\text{c}$ is the cutoff distance beyond which both the force and potential vanish. It defines a characteristic length for the DPD model and our extension, which can be used as a basis for comparison with other interaction models, particularly those that make use of a cutoff distance primarily for computational convenience rather than as a material-defining parameter. 

We generalize the conservative force in eqn (\ref{eqn:dpdf}) to arbitrary powers $n$, $m$,
\begin{equation}
\mathbf{F}^{(n,m)}_\text{c}(r_{ij})=A_{ij}\left[b_{ij}\left(1-\frac{r_{ij}}{r_\text{c}}\right)^n 
  - \left(1-\frac{r_{ij}}{r_\text{c}}\right)^m\right]\frac{\mathbf{r}_{ij}}{r_{ij}},
\end{equation}
where the repulsive exponent, $n$, represents the force order, which together with the attractive term exponent, $1 \leq m \leq n$, and two other parameters of the model, $A_{ij}$ and $b_{ij}$, completely define the interactions.
$A_{ij}$ is the repulsive parameter with the unit of force, and $b_{ij}$ 
is a scaling factor defining the magnitude of repulsion relative to attractive interactions. Here, we consider only the case of $m=1$, and in the following we omit $m$ completely for brevity, denoting the potential as $n$DPD. The corresponding energy term is, therefore,
\begin{equation} \label{eq:en}
U^{(n)}_\text{c}(r_{ij}) = \frac{A_{ij}b_{ij}r_\text{c}}{n+1}\left(1-\frac{r_{ij}}{r_\text{c}}
  \right)^{n+1}
       - \frac{A_{ij}r_\text{c}}{2}\left(1-\frac{r_{ij}}{r_\text{c}}\right)^2.
\end{equation}
From this expression it is clear that $b_{ij} > 0.5(n+1)$ is required for $U^{(n)}_\text{c}(r_{ij}) = 0$ to have a root $\sigma$ ($0<\sigma<r_\text{c}$), which from now on we identify as the nDPD particle size in a similar manner to the same parameter in the Mie and LJ potentials. Also, the energy at origin, $U^{(n)}_\text{c}(0)$, depends on both $A_{ij}$ and $b_{ij}$.
Finally, setting also $n=1$ and $b_{ij}=2$ would recover the standard DPD potential. We re-emphasise that the modification affects only the conservative interactions, preserving the Langevin part (the `DPD thermostat').

We restrict our consideration here to the first three low-order cases only, $n=2,3,4$, for a single component ($A \equiv A_{ij}$, $b \equiv b_{ij}$), leaving higher powers for subsequent publication. The cases are sketched in Fig.~\ref{fig:one}, where they are contrasted with the standard DPD potential and LJ potential (inset) used in atomistic simulations.
With the selected parameters given in Table \ref{tab:one}, the location of the first zero of the potential, $\sigma_{ij}$, is $\sim 0.5 r_\text{c}$, decreasing somewhat for higher values of $n$. By contrast, the standard DPD potential decays monotonically to zero at $r_\text{c}$. A common peculiarity of both DPD and $n$DPD potentials, distinguishing them from conventional inverse-power ones, is the relationship between the force and the potential exponents: whereas in the former the exponent in the force is lower by one from the potential power, in the latter it is higher by one.

\begin{table}
\centering
\caption{Parameters of the $n$DPD potential in reduced units. Particle size, $\sigma_{ij}$, is given as a fraction of $r_\text{c}$}
\begin{tabular}{lrrr}
$n$ & $A_{ij}$ & $b_{ij}$ & $\sigma_{ij}/r_\text{c}$ \\
\hline
 2  &  25.0    &  3.02    & 0.5033 \\
 3  &  15.0    &  7.2     & 0.4730 \\
 4  &  10.0    & 15.0     & 0.4497 \\
\hline
\end{tabular}
\label{tab:one}
\end{table}

\begin{table}
\centering
\caption{Critical Parameters of the model. Pressure and density are in scaled reduced units (see text).}
\begin{tabular}{lrrr}
$n$ & $T_\text{c}$ & $p^*_\text{c}$ & $\rho^*_\text{c}$ \\
\hline
 2  &  1.025    &  0.2951  & 0.519 \\
 3  &  1.283    &  0.3979  & 0.504 \\ 
 4  &  1.290    &  0.4095  & 0.484 \\ 
\hline
\end{tabular}
\label{tab:two}
\end{table}

\begin{figure}
\centering
\includegraphics[width=.8\linewidth]{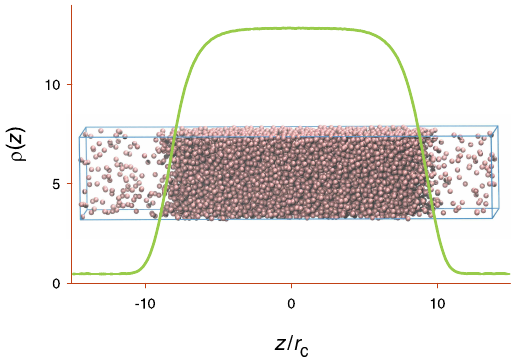}
\caption{A sketch of the system in slab geometry used to estimate the coexisting densities. The graph presents the density profile of the DPD particles (green line) giving the values of two coexisting densities, where the profile is flat, and showing two liquid-vapour interfaces at $z\approx\pm10 r_\text{c}$.} 
\label{fig:four}
\end{figure}

\subsection*{DPD simulations and units}
All $n$DPD calculations were performed using a modified version of the DL\_MESO mesoscopic simulation package\cite{Seat13,Seat21} to accommodate the new potential type. We also implemented this potential in the DL\_POLY molecular simulation package\cite{Todo06} and used it to independently verify the obtained pressure in a microcanonical ensemble (leaving Langevin forces out) in certain cases. We used DPD reduced units\cite{Alle17} throughout, with the following exception: we define the length scale in terms of $\sigma$, the shortest distance at which the interparticle potential reaches zero, as is established, e.g., for the LJ potential. We take it as the particle size, as shown in the last column in Table~\ref{tab:one}. When the properties are expressed in terms of the critical parameters (Table \ref{tab:two}), they are denoted by the asterisk superscript (e.g. $T^* = T/T_c$) and referred to as \emph{scaled reduced} properties.

Integration of the Langevin equations of motion stipulates isothermal conditions. In setting the system and during the $NVT$ simulations we used the standard velocity Verlet integrator with the DPD thermostat,\cite{Seat21} and in DL\_POLY calculations with the Nos\'e-Hoover thermostat, while for $NpT$ calculations Langevin pistons were used for both thermostat and barostat.\cite{Jako05}

Since $n$DPD interactions are smooth, bounded and of finite range, \emph{no special numerical treatment} is required to deal with their calculation for large inter-particle distances as for Lennard-Jones interactions. Such treatments can involve truncation of the potential within a cutoff distance with tapering of forces and energies to zero at the cutoff to prevent energy fluctuations and estimation of long-range tail corrections for contributions to energies and pressure.\cite{Lish18} These approaches need special care and are prone to introducing errors in calculations of thermodynamic properties\cite{Heye15} that can only be eliminated by using more accurate and expensive computational methods for long-range interactions\cite{Ewal21} to account correctly for inhomogeneous phase formation and interfacial phenomena.\cite{Klau07} 

Without the need to account for long-range effects, the only \emph{controllable} sources of systematic errors in the results are thus due to the finite timestep and the system size. Both factors were evaluated and in the simulations we used a dimensionless timestep of $0.005$. Due to the finite time step, the system temperature estimated from the ensemble-averaged kinetic energy increases quadratically from the thermostat temperature, typically by 0.1\% with the selected timestep. However, by extrapolating to an infinitesimal limit the thermostat temperature is recovered. In the fluid part of the $n$DPD phase diagram the pressure is dominated by the kinetic contribution, and finite time steps therefore lead to a quadratic rise in pressure, which amounted to an increase of one half percent near the critical point. To alleviate this effect we used the same timestep in all calculations, and for low-temperature coexistence simulations extrapolated pressure to zero timestep. With regard to the system size, we used between 8000 and 64000 particles depending on the estimated property to make the error due to system size of the order of statistical uncertainty in the results or even smaller.

\subsection*{Simulation of orthobaric densities and the solid phase}
Coexistence curves were estimated from the series of two-phase calculations 
performed in slab geometry (see Fig.~\ref{fig:four}) using rectangular elongated prism-shaped simulation boxes containing between 20000 and 64000 particles. Such a setup allows us to simultaneously estimate both the coexisting pressure and densities at a given temperature, and also the surface tension from the difference in the diagonal pressure components.\cite{Alle17}

\begin{figure}
\centering
\includegraphics[width=.8\linewidth]{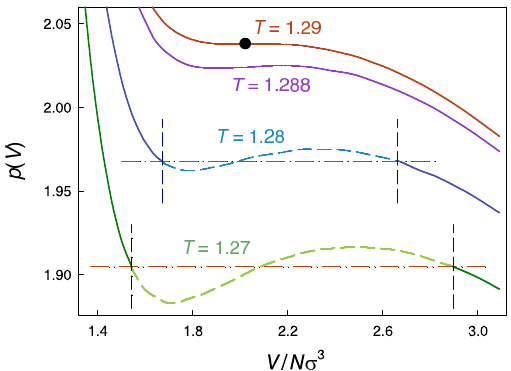}
\caption{Pressure as a function of volume per particle in terms of particle size, $\sigma$, for several near-critical isotherms for $n=4$ potential. Horizontal dash-dotted lines denote coexisting pressure from the two-phase simulation; vertical dashed lines denote the positions of coexisting volumes; black dot marks the critical point. See text for details of estimating the coexisting densities.}
\label{fig:six}
\end{figure}

Closer to the critical point the coexisting densities can no longer be reliably determined using this setup due to increased fluctuations and spontaneous creation and destruction of interfaces. The coexisting pressure can still, however, be reliably estimated as the normal pressure component along the long axis in a rectangular prism of a high aspect ratio, given that the liquid/gas interfaces have insignificant curvature. This is expected in the proximity of the critical point where the correlation length rapidly grows and the surface tension decreases. 
We estimated therefore the coexisting densities using a set of isotherms in the subcritical region, as illustrated in Fig.~\ref{fig:six}. The isotherms were constructed from a set of isochoric calculations of cubic boxes with $N=8000$ particles of progressively larger sizes. Below the critical temperature they include the van der Waals (wdW) loop as illustrated in the Figure for the $T=1.27$ and $1.28$ cases.
Although for finite volumes all states along the isotherms are thermodynamically stable,\cite{Bind12} the regions marked by the dashed lines in two lower curves may or may not phase separate inside the boxes.\cite{Bind12} These regions, of course, are metastable or completely unstable in the thermodynamic limit, which precludes us from using the Maxwell construction.\cite{Maxw75}
Instead, we used the pressure values obtained in two-phase simulations (the horizontal dashed lines in the Figure) and estimated the coexisting densities from the crossing points with the isotherm where it has a negative slope, as marked by the vertical dashed lines in the Figure. In this way our estimates are based only on thermodynamically safe data from the regions covered by the full lines in the Figure and allowed us to obtain the densities for temperatures as close as $T=0.99T_\text{c}$.

We determined the type of the solid structure by annealing the systems, each containing $4N^3$ and $2N^3$ particles (where $N$ is an integer, we used between 12 and 20) and thus promoting formation of BCC and FCC solids correspondingly, and allowing the solid phases to form. We subsequently used common neighbour analysis (CNA), as implemented in the Open Visualization Tool OVITO,\cite{Stuk09} to analyse the particle trajectories and determine the preferred type of the cubic cell. The ratio of FCC to BCC local structures was more than 20 in all cases, from which we concluded that the equilibrium structure is FCC. 

\begin{figure}
\centering
\includegraphics[width=.94\linewidth]{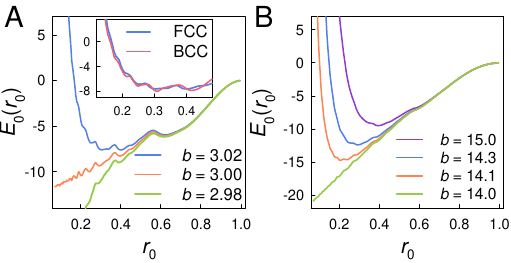}
\caption{Zero-temperature energy per particle as a function of interparticle nearest-neighbour separation. (\emph{A}) FCC lattice with $n$DPD potential $n=2$ for three values of the repulsion parameter; inset - comparison of FCC and BCC energies for $b=3.02$. (\emph{B}) energies for $n=4$ potential for four values of the repulsion parameter.}
\label{fig:five}
\end{figure}

In order to determine the ground state (zero temperature) energy and structure we calculated the energy of the FCC lattice as a function of the nearest-neighbour distance $r_0$, the results are shown in Fig.~\ref{fig:five}. Thermodynamic stability requires the condition that the total configurational energy of a system of $N$ particles is bounded from below\cite{Fish66}, i.e., that $U_N(r^N)\geq -NB$, where $B$ is a fixed number. This condition is controlled by the $n$DPD parameter $b_{ij}$, which for a single-component system with this pairwise potential resolves to\cite{Loui00}
\begin{equation}
b > \frac{(n+1)(n+2)(n+3)(n+4)}{120},\label{eqn:ndpdstable}
\end{equation}
and the values in Table \ref{tab:one} were selected to satisfy it. The limiting values of $b$ in the 2DPD ($n = 2$), 3DPD ($n = 3$) and 4DPD ($n = 4$) cases are thus 3, 7 and 14 respectively.

The oscillations in energy, well resolved for the $n=2$ potential (Fig.~\ref{fig:five} A), result from new members of the lattice entering the range of attractive potential forces as the system compresses. The oscillations indicate an intriguing possibility of different polymorphs in the solid state, something to be looked into in the future. The inset shows the energy per particle density dependence in the FCC and BCC lattices, indicating a series of FCC$\leftrightarrow$BCC transitions as the pressure increases. Choosing a parameter $b$ below the stability limit will lead to a system collapse akin to gravitational collapse in neutron stars, as seen from the cases of $b=2.98$ for 2DPD (Fig.~\ref{fig:five}A) and $b=14.0$ for 4DPD (Fig.~\ref{fig:five}B); it is also supported by our simulation tests.

\subsection*{Critical point and critical constants estimation}
We determined the location of the critical point from a set of isotherms at progressively increased temperatures. The first isotherm that does not contain the vdW loop, here the $T=1.29$ isotherm, is deemed to be the critical one.
To estimate the critical density of the model we fit the Wegner expansion,\cite{Wegn72} $\rho_\text{l}(\tau)-\rho_\text{v}(\tau)=A\tau^{\beta} + B\tau^{\beta+\Delta}$, where $\tau=1-T/T_\text{c}$, to our coexistence data, with $\beta=0.32653$ and $\Delta=0.5$ assuming the commonly accepted 3D Ising universality class\cite{Wegn76} and using the fact that the lattice gas model and the Ising model are isomorphic.\cite{Lee52}

\subsection*{Solid--liquid phase transition}
In order to locate the solid--liquid transition control parameters (transition temperature in the $T$-scan and transition pressure for pressure-induced melting) we initially established the approximate area of transition by scanning the control parameters and monitoring the density (volume) for a system of $N=32000$ particles. This gave us roughly $\pm 20$\% margins in the control parameters around the transition.
To solve the nucleation problem, we then prepared an initial two-phase state by equilibrating separately bulk liquid and bulk solid samples using the parameters outside of the transition margins and then bringing the two systems in contact, watching for the absence of significant overlaps at the flat boundary. Using a sample of $N=64000$ particles with one half in a typical liquid configuration and the other in a typical solid one, we conducted a series of short $NpT$ simulations at progressively higher values of the control parameter, monitoring the kinetics of the interphase boundary. The direction of the boundary motion (towards freezing or melting of the whole sample) appears to be a sensitive indicator of the ensuing phase, insensible to the initial configurations, and we were able to locate the transition within a $\pm 1\%$ margin for the temperature and within $\pm 10\%$ for the pressure.

\section*{Results}
\subsection*{Properties along the liquid-gas coexistence curve}
\begin{figure}
\centering
\includegraphics[width=.8\linewidth]{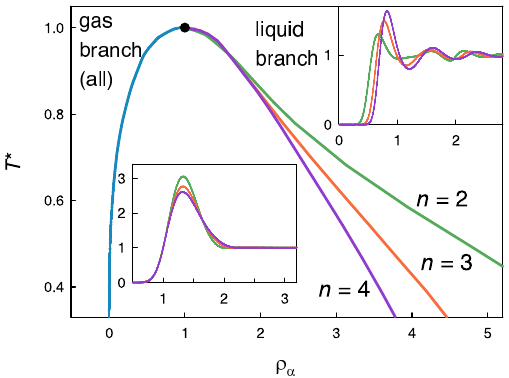}
\caption{Coexistence curves for the $n$DPD potential, calculated for quadratic, cubic, and quartic repulsive forces with parameters given in Table \ref{tab:one}. The blue line denotes the gas branch (common for all powers) and the green, orange, and red lines are for the liquid branches. The black dot denotes the critical point ($T^*=1$). Reduced units have been used with critical values defining the scales. (\emph{Insets}) Radial distribution functions at the coexistence densities: left - gas branch, right - liquid branch, at $T^*=0.4$ for the three potential classes.}\label{fig:two}
\end{figure}

Using the $n$DPD potential with $n=2,3,4$, we explored both branches of the binodal of their fluid phases from the triple point to the critical point. The binodal curves presented in Fig.~\ref{fig:two} in scaled reduced units (with respect to the corresponding critical density and critical temperature, see Table \ref{tab:two}) demonstrate several unusual features. First, the liquid phase in the $n$DPD model has a much wider extent than simple atomic liquids and  water. For example, for the $n=4$ (4DPD) potential the liquid phase extends down to  $T^*=0.0709$ ($T=0.0915$), whereas argon has its triple point at $T^*=0.556$, and water at $T^*=0.422$.
Second, while the gas branch in all three cases falls on the same curve on the scale of the figure, the liquid branches are markedly different outside of the scaling region. Even the concavity of the liquid branch varies with the parameter $b$ of the potential: with the selected parameters the liquid branch is concave for $n=4$, a straight line for $n=3$, and a convex curve for $n=2$, challenging the law of corresponding states.
We have verified that the $n=2$ branch becomes concave when $b$ increases from 3.02 to 3.1. The straight line appears to be the limiting case for $n\geq 3$. 
In the scaling region near the critical point all liquid branches collapse onto a single universal curve. 

The insets to the Figure illustrate the radial distribution functions (RDF) for both branches at a representative temperature. For the gas branch all cases display a typical low-density atomic (e.g., LJ fluid)  RDF, with its height increasing with the potential well depth: more so for lower powers of potential ($n$). Conversely, the liquid RDFs show a transition from the usual simple fluid liquid-density RDFs for the two higher powers to an unusual fluid for the 2DPD potential with the second minimum more pronounced than the first, where the positions of the first and the third solvation shells are also shifted to shorter distances but the position of the second one remains unchanged. We have found these features of the liquid-density RDF to be sensitive to the value of $b$ -- they disappear when $b$ is increased from 3.02 to 3.10 -- and are not unique to 2DPD: the second minimum becomes similarly pronounced when $b < 7$ for 3DPD and $b < 14.5$ for 4DPD.

\begin{figure}
\centering
\includegraphics[width=.94\linewidth]{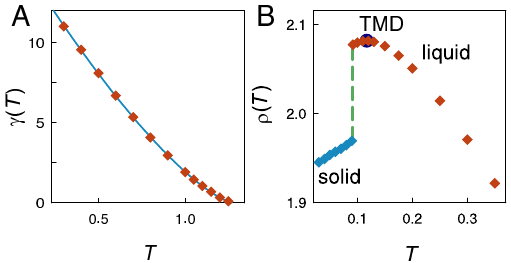}
\caption{(\emph{A}) Calculated surface tension of 4DPD liquid in DPD units as a function of temperature up to the critical point. Solid line is a power law fit (see text). (\emph{B}) Temperature dependence of density around the solid-liquid phase transition. Red diamonds - liquid phase, blue diamonds - solid phase, and the temperature of transition is marked by the green dashed line. The large blue dot indicate the position of the temperature of maximum density (TMD).}
\label{fig:three}
\end{figure}

We calculated the surface tension $\gamma$ as a function of temperature and, by fitting the data to a power law $\gamma\propto(T_\text{c}-T)^\mu$, we obtained an independent estimate of the critical temperature. For the 4DPD potential, illustrated in Fig.~\ref{fig:three}\emph{A}, $T_\text{c}=1.28(2)$, in excellent accordance with the value obtained from the binodal data ($T_\text{c}=1.29$). The second parameter of the fit, the critical exponent $\mu$, has a converged value of 1.38, deviating somewhat from the accepted value for fluids, $\mu=1.28(6)$.\cite{Wido72} In the Figure we expressed the surface tension in the standard DPD units (length in units of potential range $a$). This facilitates the comparison with experimental data for key solvents. For example, water modelled as a DPD fluid, with an $a=8.52$~\AA{} (7:1) mapping would have a surface tension of $\gamma=12.7$ at $T^*=0.46$ (based on 72.0 mN~m$^{-1}$ at $T=298$~K.\cite{Pall90})

\subsection*{Contraction on fusion, temperature of maximum density, and negative thermal expansion in the solid}
At low temperatures, an $n$DPD fluid freezes for all three studied classes, which is where it exposes another unusual behaviour. Whereas most substances expand when heated, there is a compelling group of materials that contract upon heating within a certain temperature range. Classic textbook examples include liquid water below 4 $^{\circ}$C (277 K) and ice below 63 K.\cite{Dant62} This phenomenon, termed Negative Thermal Expansion (NTE),\cite{Dove16} occurs when materials contract in volume upon heating and is characterised by the negative value of the bulk thermal expansion coefficient, $\alpha=(\partial\ln V/\partial T)_p < 0$. NTE in complex materials with directional bonding, where it is more often found,\cite{Pryd96,Dove16} received much attention in recent years due to its industrial applications. Molecular mechanisms explaining NTE in these systems include, inter alia, Gr{\"u}neisen vibrational theory of thermal expansion,\cite{Barr05,Dove16} special phonon properties,\cite{Pryd96} and rigid unit modes.\cite{Pryd96} At the same time, there are a few pure elements where the NTE and contraction of fusion have been found, neither of which can be explained by the above mechanisms. These elements include the semi-metal bismuth, the metals gallium and plutonium, 
and the metalloids antimony, germanium, and silicon. They crystallise into different crystal structures,\cite{Young91} belong to different groups in the periodic table, and exhibit markedly different NTE ranges: it is thus unlikely that their behaviour can be attributed to unique electronic properties alone. There is no clear understanding of the underlying mechanisms in these systems.
Furthermore, several classical single-component model systems with isotropic pair potentials, such as the purely repulsive Gaussian core model,\cite{Stil76,Stil97} as well as the ad hoc devised potentials with a softened interior of their basins of attraction,\cite{Rech07} also feature NTE. However, finding a simple model with an isotropic potential, which reproduces the anomalous behaviour of water\textemdash{}contraction on melting with an immediate negative expansion region in the liquid phase\textemdash{}is known to be a challenging open problem.\cite{Rech07} Here, we demonstrate that the $n$DPD model includes all of the essential ingredients required to capture this complex phenomenon.

In order to locate the solid-liquid phase transition we calculated the isobaric density of $n$DPD at low temperatures.
The results for the $n=4$ potential at zero external pressure are presented in Fig.~\ref{fig:three}\emph{B}. The freezing transition occurs at $T=0.0779$ ($T^{*}=0.0760$) for the 2DPD potential, at $T=0.0873$ ($T^{*}=0.0680$) for the 3DPD potential, and at $T=0.0915$ ($T^{*}=0.0709$) for the 4DPD potential. The system freezes into an FCC solid with an expansion in volume of ca 5.5\% in all three cases. Just before freezing a density maximum is observed at $T=0.11$ ($T^{*}=0.0852$) for the 4DPD case. On the liquid side, it qualitatively describes water, which has a temperature of maximum density (TMD) of 4 $^{\circ}$C (277 K) or, in reduced units, $T^{*} \approx 0.4283$. Even taking into account that water expands by 9\% upon freezing, $n$DPD presents alternatives to hydrogen bonding mechanisms that lead to the qualitatively same effect, which could be explored further when developing accurate CG water models.

\subsection*{Pressure-induced melting}
If the liquid phase of a substance is more dense than its solid phase, the melting temperature decreases with increased pressure. This general result, which follows from the Clausius-Clapeyron equation, is well known for water, where a regelation occurs for up to 20~K below the normal freezing point. A similar phenomenon is also observed for the chemical elements mentioned in the previous section that feature NTE, as well as in alkali metals. Using the two-phase setup to study the solid-liquid transitions, we observed a monotonous decrease in transition temperature from $T^*=0.0709$ at effectively zero pressure down to $T^*\approx0.03$ at $p=10$ for a 4DPD potential. As a consequence of this, there is a maximum temperature for which the solid phase exists for the  $n$DPD model: $T^*=0.0709$ for 4DPD. For temperatures below this maximum value, the system melts when pressurised. It should be noted that other models also feature pressure-induced melting. For example, in the Gaussian core model\cite{Stil76} reentrant melting has been reported.\cite{Giaq05}

\subsection*{Parameterisation of $n$DPD and thermodynamic consistency}
In order to relate thermodynamic properties at CG and atomistic levels a mapping between corresponding units is needed. At equilibrium, this includes at least energy and length scales.
In the DPD model the link to a specific chemistry requires a mean-field approximation.\cite{Groo97} With only one parameter in the model, $A_{ij}$, it is sufficient to fix it by matching the isothermal compressibility of DPD fluid and that of a given material, e.g. water.\cite{Groo97} This is justified by the fundamental connection between DPD and fluctuating Landau-Lifshitz Navier-Stokes hydrodynamics.\cite{Espa03}

The elastic responses of a $n$DPD liquid are richer and also include tensile stresses with the tensile strength controlled by the parameter $b_{ij}$. In addition, the liquid-gas coexistence present in the $n$DPD model with its associated critical point can fix the temperature scale. Thus, $A_{ij}$ can be fixed from the condition $T_\text{c}=1$, and $b_{ij}$ by mapping the surface tension of the $n$DPD model and the target fluid. The value of $b_{ij}$ is limited by the need for thermodynamic stability\cite{Fish66,Loui00}, which is satisfied for a given value of $n$ by eqn (\ref{eqn:ndpdstable}). 

The length scale for the $n$DPD model can be fixed by relating liquid coexisting densities between simulation and physical representations. Although we did not aim for a specific system here, a relation to ambient water can be made as follows. The length scale in DPD is usually fixed by mapping 3 water molecules to one bead.\cite{Groo01} There is no compelling reason for this choice, and due to tetrahedral ordering of liquid water at ambient conditions a 5:1 mapping, i.e., a bead representing a water molecule together with its first coordination shell, seems equally sensible. Since the liquid density of water is $1/30$~\AA$^{-3}$,\cite{Groo01} and the corresponding 4DPD density is $1.62\sigma^{-3}$, both taken at $0.4608T_\text{c}$ (298~K), the length scale using this mapping is $\sigma=6.24$~\AA, and the cutoff distance $r_\text{c}=13.88$~\AA.

\begin{figure}
\centering
\includegraphics[width=.94\linewidth]{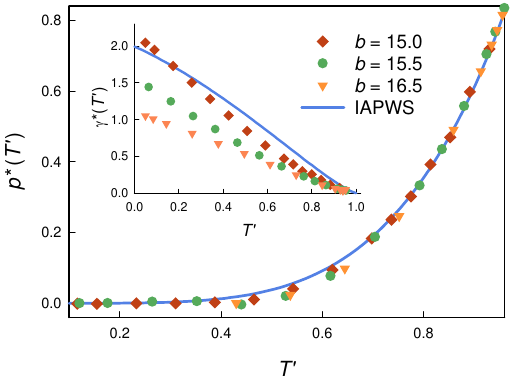}
\caption{Saturation pressure in scaled reduced units of a 4DPD liquid with three values of the $b$ parameter (symbols), and the water using the reference IAPWS equation of state\cite{Wagn02} (full line). The standard error in our data is of the order of symbol size. In the inset, the surface tension for these three cases is compared with the IAPWS reference.
The temperature scales from the triple point for IAPWS or zero pressure melting point for 4DPD ($T'=0$), to the critical temperature ($T'=1$).}
\label{fig:saturation}
\end{figure}

With these parameters we can link the liquid phases of water and $n$DPD. Most conveniently, this could be done using a linear transformation $T'=(T-T_0)/(T_\text{c}-T_0)$, where the scaled temperature $T'$ is set to zero at the zero-pressure melting point (triple point for water), $T_0$, and increases linearly with $T$ reaching a value of 1 at the critical temperature $T_\text{c}$. We have found that a 4DPD liquid with a $b$ parameter in the range $15.0\leq b\leq 16.5$ accurately describes the saturation pressure of water across a wide range of temperatures, as illustrated in Fig.~\ref{fig:saturation} with a comparison between our results and the reference equation of state given by the International Association for the Properties of Water and Steam (IAPWS).\cite{Wagn02}

The surface tension of the $n$DPD liquid is obtained as a by-product from the difference between normal and lateral pressure components in a coexistence simulation using a slab geometry.\cite{Alle17} Since the $n$DPD potential is of finite range, no long-range corrections are required. An inset to Fig.~\ref{fig:saturation} shows the temperature dependence of the surface tension of the 4DPD case for three values of $b$. The surface tension, given in the Figure in DPD scaled units ($T_\text{c}=1$), markedly depends on $b$, which provides a suitable method of tuning it to a specific target. Although the surface tension is not reproduced over the entire temperature range by a single set of parameters, it can be tuned around a particular temperature. Thus, $b=15.0$ seems a reasonable choice for ambient water, $T'=0.0668$ (298~K).

\section*{Discussion and Conclusions}
In his milestone paper discussing the hierarchical structure of science, `More is different',  Phillip Anderson argued that behaviour of complex systems cannot be entirely `understood in terms of a simple extrapolation of the properties of a few particles', and that `at each level of complexity entirely new properties appear'.\cite{Ande72} As we dial down the magnification moving to larger scales, we would find the new phenomena in focus, and these phenomena might or might not be present at finer resolutions. 
Liquid structure is defined by correlations between the density fluctuations, which includes both short-range order and long-range hydrodynamic fluctuations. However, short-range and long-range hydrodynamic fluctuations are coupled, and coarse-graining the short-range interactions only will inevitably disrupt the balance. In designing a coarse-grained model there are a couple of specific points that needs to be taken into account.

It is well-known that coarse-graining of any set of degrees of freedom results in changes to the governing equations of motion from the Newtonian form to generalised Langevin dynamics (GLE),\cite{Zwan01} i.e., to the dynamics of interacting Brownian particles. Thus, the structure and dynamics of a CG fluid are coupled: a part of the (conservative) \emph{chemically-specific} interatomic interactions, not captured by the inter-particle interactions at the CG level, is reinstated in the form of non-conservative \emph{generic} interactions with the `bath', i.e., in the form of viscous and stochastic forces coupled via the fluctuation-dissipation theorem.\cite{VanK07} We note in passing that, strictly, even atomic-level dynamics are also of the GLE type since empirical interatomic potentials effectively include averaged-out electronic degrees of freedom. While the neglected forces are likely to be insignificant at the atomic level, this is not so at the mesoscopic scale. Incorporating memory effects is a challenging problem and is the focus of current research,\cite{Hijo10,Klip21} although no viable scheme for integration of GLE exists as yet. Existing models are usually based on a Markovian (memoryless) approximation, which enormously simplifies the equations of motion but requires \emph{effective} (mean-field) friction coefficients though obtainable, as has been demonstrated recently,\cite{Sokh19} by direct ensemble averaging of an atomistic simulation. In the CG process it is the short-range correlations that are suffer most.

Another moot point is related to generating adequate representations of many-body interactions using pairwise CG potentials. In other words, whether it is possible to retain long-range correlations in liquid with depleted short-range ones, since they are related via the Ornstein-Zernike (OZ) equation?\cite{Rowl82} In general, any adjustment in the Hamiltonian alters the thermodynamic phase behaviour.\cite{Cous95} This is true even if the CG Hamiltonian is obtained using the rigorous Mori-Zwanzig projection operator formalism\cite{Zwan01}\textemdash{}the intrinsic reduction in entropy will affect the thermodynamic functions of the CG system.\cite{Dunn16} With the reduced set of \emph{effective} degrees of freedom certain correlations are lost and, with them, all related phenomena can be missed. 
This problem of \emph{representability}, which has been often skirted in previous CG studies, is now receiving much attention.\cite{Dunn16,Noid13,Wagn16}
The development of rigorous bottom-up approaches with chemical accuracy is further hampered by the limited transferability of CG models and economic factors, which need to be taken into account: honing the CG model to specific chemistry may require more effort than solving the problem in question at the atomistic level.
At the same time, there is always great demand for finding models that would provide greatly simplified points of view, and as such there is a continuous quest for simple generic models with extended representability. Simple intermolecular interactions do not necessarily entail simple thermodynamics,\cite{Male07} and analytically simple soft potentials may conceal unexpected macroscopic phenomena.\cite{Stil76,Teje94,Pres14,Jagla98}
In order to shed more light on the impact of CG of a system on its thermodynamics we study the phase diagram of a simple soft pair potential that is a generalization of the standard potential used in Dissipative Particle Dynamics (DPD). 


Despite the wide popularity of the DPD model,\cite{Espa17} little justification has been provided for its harmonic repulsion potential.\cite{Groo97} It might therefore be surprising that a simple repulsive force of the DPD model can provide a working solution for many cases.\cite{Espa17} To qualitatively understand this success, imagine a coarse-graining process with progressively larger scaling factors. The liquid structure is defined in terms of density correlations, which includes\textemdash{}according to the OZ approach\cite{Rowl82}\textemdash{}a short-ranged component referred to as a direct correlation function. In the OZ picture the total correlation between two particles is composed of the direct correlation and an indirect part, which is longer-ranged and composed of all possible particle chains of direct correlation in the fluid.\cite{Rowl82} In the CG process the direct correlations are averaged out, which leaves just indirect correlations to be reproduced by a suitable ansatz. At this level of CG the interactions are much softer, and we argue that they are captured by the $n$DPD model. Since much of the chemical specificity is absorbed within the short range, the longer-range correlations are of a more generic nature. At even higher CG scaling most of the correlation is averaged out, leaving only the soft elastic repulsion between the beads due to the finite compressibility of the fluid. The interactions emerging at this scale are therefore of the DPD type.

As we have demonstrated, the $n$DPD potential with appropriate choices for its three parameters is able to produce stable liquid and solid phases. At the same time, the phase diagram of $n$DPD matter is topologically different from that of simple atomic systems, with the solid phase occupying a narrow density region at low temperatures. The thermodynamics of its condensed phases are rich with anomalous features, usually attributed to complex molecular systems. They include expansion on freezing, temperature of maximum density, negative thermal expansion in the solid phase, an unusual liquid-vapour phase envelope and pressure-induced melting. In this work, we have not explored in full the model parameter space, particularly the relative range of the attraction, which is likely to conceal additional anomalies. However, the fact that the anomalies are found in systems with soft \emph{coarse-grained} isotropic pair potentials indicates that they likely to originate in the medium-range structure, i.e., beyond the first neighbours in the underlying atomistic system. The complex phase behaviours have also been found in models with coarse-grained bonded interactions\cite{Teje94,Male07} and were observed in colloidal suspensions.\cite{Puse86}

Clearly, no CG model can reproduce the whole spectrum of observables of the underlying atomistic model, in the same way as no atomistic model can reproduce the entire set of experimental observables. This is a manifestation of the representability problem,\cite{Wagn16} which arises due to the resolution change at the mean-field level: certain correlations are missing. Thus, a CG model cannot describe the processes of self-assembly or gelation, which involve changes in fine-level molecular topology. However, this is not the purpose of the model. Rather, each model level in the hierarchy defines its own classes of \emph{compatible} observables that reproduce the experimental observables within the model constraints.\cite{Wagn16} A good model goes beyond this and has predictive power. Paraphrasing the famous designer Yves Saint-Laurent, one can say that a good model can advance the field by ten years.\cite{Bail84}


\section*{Conflicts of interest}
There are no conflicts to declare.

\section*{Acknowledgements}
This work was supported by the STFC's Innovation and Development Programme (IDP) and made use of ARCHER2 computational support by CoSeC, the Computational Science Centre for Research Communities, through the High-End Computing Materials Chemistry Consortium (MCC).  Additional computing resources were provided by STFC Scientific Computing Department’s SCARF cluster.  Contributing funding was also provided via the Virtual Materials Marketplace (VIMMP) project from the European Union’s Horizon 2020 Research and Innovation Programme (Grant Agreement no. 760907).


\balance


\providecommand*{\mcitethebibliography}{\thebibliography}
\csname @ifundefined\endcsname{endmcitethebibliography}
{\let\endmcitethebibliography\endthebibliography}{}

\end{document}